# TOWARD A COMPARATIVE COGNITIVE HISTORY: ARCHIMEDES AND D. H. J. POLYMATH

Lav R. Varshney

IBM Thomas J. Watson Research Center
19 Skyline Drive
Hawthorne, New York 10532, USA
e-mail: lrvarshn@us.ibm.com

## ABSTRACT

Is collective intelligence just individual intelligence writ large, or are there fundamental differences? This position paper argues that a cognitive history methodology can shed light into the nature of collective intelligence and its differences from individual intelligence. To advance this proposed area of research, a small case study on the structure of argument and proof is presented. Quantitative metrics from network science are used to compare the artifacts of deduction from two sources. The first is the work of Archimedes of Syracuse, putatively an individual, and of other ancient Greek mathematicians. The second is work of the Polymath Project, a massively collaborative mathematics project that used blog posts and comments to prove new results in combinatorics.

## INTRODUCTION

The Greek geometers of antiquity such as Euclid of Alexandria, Archimedes of Syracuse, and Apollonius of Pergaeus lived in a world very different from today. There was perhaps only one mathematician born per year, spread rather thinly across the eastern Mediterranean. Ancient Greek mathematicians were isolated both in space and in time (Netz, 1999, p. 291). Moreover, their repertoire of technological tools did not include pen/paper, chalk/blackboard, or modern information technologies, but was limited to wetted sand and stick (Netz, 1999, p. 14). Nevertheless, the mathematical results that they produced are well-recognized as the products of intelligence and are often said to have laid down the deductive tradition of modern mathematics.

In the modern world occupied by Fields Medalists such as Terence Tao and W. Timothy Gowers, mathematics has become central, with thousands of mathematicians around the globe connected by information technologies that allow communication faster than the blink of an eye. Large, loosely organized groups of people, whether mathematicians or otherwise, can now work together electronically in effective ways. Groups of individuals collectively doing things that seem intelligent has been termed *collective intelligence* (Malone et al., 2010). Mathematical proof, a kind of problem-solving (Hong and Page, 2001), is one exemplar of intelligent things.

Previous studies of collective intelligence (Malone et al., 2010) have developed a taxonomy organized around building blocks of:
- What is being done?,
- Who is doing it?,
- Why are they doing it?, and
- How is it being done?

Although such a taxonomy broadly defines how collective intelligence is performed, there is insight to be derived from a detailed focus on the 'doing of intelligence' in this new form.

Woolley et al. (2010) have studied collective intelligence from a cognitive science perspective. They define a group's collective intelligence to be the general ability of a group to perform a wide variety of tasks, and think of it as a single scalar quantity. They find that collective intelligence is not just the average of individual intelligences, but there are other factors at play: it is a property of the group itself, not just the individuals comprising it.

In particular, they find that collective intelligence is correlated with network properties such as the average social sensitivity of group members, the equality in distribution of conversational turn-taking, and the proportion of females in the group.

Rather than determining that collective intelligence can be measured and which features other than constituent individual intelligences are correlated to it, our goal here is to see if there are differences in the process of cognition: *is collective intelligence different in kind from individual intelligence?*

Many have argued for a modular theory of mind (Fodor, 1983), with a brain composed of many agents (Livnat and Pippenger, 2006). Indeed, if the individual mind is modular, then the processes of cognition in individual intelligence may not be too different from those in collective intelligence.

As it turns out, individual intelligence has been correlated with network properties of the brain, just as collective intelligence has been correlated with network properties of the group. Efficient information flow correlates positively with intelligence, since high intelligence probably requires smooth information transfer among brain regions. Moreover, intelligent brains process information more efficiently by using fewer brain resources when performing cognitive tasks (Deary et al., 2010).

To study whether collective intelligence is different from individual intelligence, we take the position that artifacts from the process of cognition should be examined and compared quantitatively. Historical artifacts are a product of a given place and time, and may be thought of as limited, but they provide a kind of detailed description and insight that even controlled experiments in cognitive science cannot.

As such, methods from *cognitive history*, i.e. introduced by Netz (1999), should be adopted. Moreover, since two different kinds of artifacts are to be compared, the result will be a comparative study that in some sense follows Mill's canonical inductive inference methods based on agreement and difference (Mill, 1872). We term our approach *comparative cognitive history*.

The artifacts of mathematical deduction and proof serve as a good starting point for this research agenda since they are discrete objects whose structures are easily captured and quantified.[1]

We consider the two following examples of artifacts of deduction for our case study. For individual intelligence, we consider the geometry theorems of ancient Greek mathematicians such as those in Euclid's *Elements*, Archimedes' *The Method*, and Apollonius' *Conics*. We examine the structure of argument that they use in their proofs (Netz, 1999), treating this structure as a directed graph. For collective intelligence, we consider the combinatorial proof of the Density Hales-Jewett theorem that was developed by the Polymath Project (Cranshaw and Kittur, 2011). Rather than looking at the proof itself, we look at the structure of arguments that were put forth in developing the proof. Again, this structure can be treated as a directed graph.

To compare directed graphs, we adopt measures from network science (Jackson, 2010; Easley and Kleinberg, 2010) including degree distributions and subgraph distributions. Subgraph distributions have been used to define structural motifs (Milo et al., 2002) and families of networks (Milo et al., 2004).

We see that the structure of the Polymath Project work involves much broader degree distributions than the work of say Archimedes. That is, in the Polymath Project there are instances where a large number of statements are used to build an argument and instances where a single statement is used to support many arguments. We also see that the 'Aristotelian syllogism' subgraph appears more frequently in Archimedes than in Polymath, suggesting that this was a commonly used argument structure in Greek antiquity. Moreover, simple implications are the predominant form of argument in Polymath, suggesting a need to keep things simple.

The remainder of this position paper is organized as follows. First, we provide background on methods of cognitive history and quantitative epistemology and argue for the use of a quantitative comparative cognitive history to study individual and collective intelligence. Next we initiate a case study to demonstrate the validity of the proposed research agenda, by detailing the background of two artifacts. Quantitative characterization and comparison is presented next. Finally, limitations of the current case study and next steps for the broader program of studying the 'doing of intelligence' are given.

## **COGNITIVE HISTORY**

The methodology of cognitive history was introduced by Netz (1999) to study the nature of deduction. As others have noted, however, the basic methodology is common to other parts of science studies with its "obsessive attention to the material, historical, and practical conditions necessary for the discovery of new cognitive skills" (Latour, 2008).

As its name implies, cognitive history lies at the intersection of cognitive science and history. Like cognitive science, it approaches knowledge through its structural forms and practices rather than its specific propositional content. Like the history of science, it studies specific cultural artifacts rather than through experiments whose results are meant to be universally applicable.

---

[1] Note that although there are many social and epistemic considerations surrounding the nature of deduction, see e.g. (MacKenzie, 2001), we do not deal with them here.

By focusing on specific historical artifacts rather than potentially generalizable phenomena as in cognitive science, one might wonder what is lost. Looking at it another way, even though there are generalizable measures of intelligence (Deary et al., 2010; Wooley et al., 2010), it is not clear that there are universal rules that govern the process of reasoning.

While there may be no general, universal rules for reasoning, such rules do exist historically in specific contexts (Netz, 1999, p. 6). Reasoning is done in a very specific way and there is a method to the deployment of cognitive resources. Specific cognitive methods are specific ways of 'doing the cognitive thing.' They change over historical timescales and so may seem relatively stable for long periods, but are not constant (Netz, 1999). Technological change, including the introduction of new information technologies, changes cognitive processes (Sparrow et al., 2011).

Hence cognitive methods should be studied as historical phenomena, valid for their period and place, rather than universally so. Our comparative cognitive history approach takes this position.

Netz (1999) used some basic quantitative methods in his study of deduction in Greek mathematics, but our position is that this can be taken further by adopting methods from network science and discrete mathematics. In this sense, we follow ideas from *quantitative epistemology*.

Quantitative epistemology studies the nature of knowledge with numerical techniques. It is said that "such techniques enable one to supplement intuitive or impressionistic analyses by re-framing qualitative problems in quantitative terms" (Barany, 2009). Quantitative epistemological analysis proceeds in three steps (Barany, 2009):
1. Capture knowledge in a form that can be analyzed quantitatively,
2. Develop means of quantification to match epistemic intuitions, and
3. Use mathematical techniques to study these quantifications

so as to aid in understanding the systems of knowledge under consideration. We use this basic checklist in our case study to demonstrate its validity for comparing the structure of knowledge arising from examples of individual and of collective intelligence.

## BACKGROUND ON ARTIFACTS

### D. H. J. Polymath
The aim of the Polymath Project was twofold: first to find an elementary proof of a special case of the Density Hales-Jewett (DHJ) theorem, which is a result in combinatorics that can be interpreted in terms of playing multidimensional tic-tac-toe; and second to demonstrate the utility of collective intelligence for mathematical research (Gowers and Nielsen, 2009).

Participation was open to anyone in the world, and the primary approach taken was the development of small statements and arguments towards a proof of the theorem. The main venue for participants to work was within the posts and comments of the personal blogs of noted mathematicians Timothy Gowers and Terence Tao. Rules explicitly discouraged participants from working extensively on their own without discussing progress on the blog (Cranshaw and Kittur, 2011).

In initiating the Polymath Project, "Gowers sought the kind of free-wheeling conversational interplay one finds in interpersonal mathematical collaboration at its best" (Barany, 2010). Blogs are an appropriate technology since they have a temporal ordering and organization, just like conversations (Barany, 2010).

Although a small fraction of contributors created most of the content—notably Gowers and Tao—almost all contributors provided some content that was influential to the task of deduction (Cranshaw and Kittur, 2011).

The end result was indeed a combinatorial proof of the DHJ theorem, which has been submitted for publication under the nom de plume D. H. J. Polymath (Gowers, 2010). More importantly for our purposes, however, the numbered blog comments and their reference structure left a historical record of the process of deduction by a collective intelligence. We use the structure of these comments as reconstructed by Cranshaw and Kittur (2011).

Tao suggested that in its broader principles there was little to differentiate Polymath from ordinary mathematical research (Barany, 2010), however it is our position that this is an open question to be studied.

### Archimedes (and Euclid and Apollonius)
In looking at artifacts of deduction that arise from individual intelligence, we consider a few proofs of theorems from Greek antiquity. In particular:
- Euclid's *Elements* II.5

- Archimedes' *The Method* I
- Archimedes' *On Sphere and Cylinder* 1.30
- Archimedes' *Spiral Lines* 9
- Apollonius' *Conics* 1.41

These works are considered classics in the history of rigorous deduction, focusing on problems in geometry.

Greek mathematics was often reported in exchanges with other potentially interested parties. These mathematical exchanges were accompanied by lettered diagrams that defined the closed deductive system. An exclusively oral presentation was therefore ruled out, leaving the fully written form for addressing mathematicians abroad and a semi-oral form, with some diagram, for presentation to a small group of fellow local mathematicians (Netz, 1999, p. 14). Greek mathematics was like a conversation.

In coming forward to the modern era, theorems and proofs have been carried through various intermediary people and technologies (Netz and Noel, 2007), however they retain the characteristics of individual intelligence. We use the structure of these proofs as reconstructed by Netz (1999). This consists of an ordered sequence of statements and references to previous statements that are used to make arguments in support; there are also starting-point statements that require no justification.

It is said that "Greek mathematical proofs are the result of genuine cross-fertilization," (Netz, 1999, p. 170) involving the use of many different sources of starting-points. Indeed, "derivatives of a single assertion must carry similar informative contents, whose intersection could be neither surprising nor revealing," (Netz, 1999, p. 170) and a desire of many Greek mathematicians was to be playful and surprising (Netz, 2009).

## **STRUCTURE OF ARGUMENT AND PROOF**

Having described the artifacts under study, we now proceed to demonstrate our quantitative cognitive history program. The first step is a quantitative representation of the artifacts.

In discussing the deductive process in the Polymath Project, we use the notion of an *argument graph*. Such a graph has nodes that correspond to numbered statements and directed edges that correspond to references among statements. The statements are comments on blog posts. Due to the time-ordering of comments, the graph is a directed acyclic graph. Although it has many connected components—often comments that did not lead anywhere—we look at the largest component with 299 comments that is shown in Figure 1.

In discussing the logical process represented in the proofs of Archimedes and others, we also use argument graphs. Nodes are statements and directed edges are from a statement that is supported by another statement through citation. Due to the logical flow of proofs, the graphs are acyclic. All graphs are a single component. Figures 2–6 show these graphs.

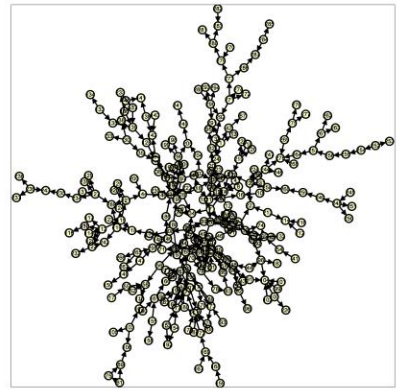

*Figure 1:* Polymath Project argument graph.

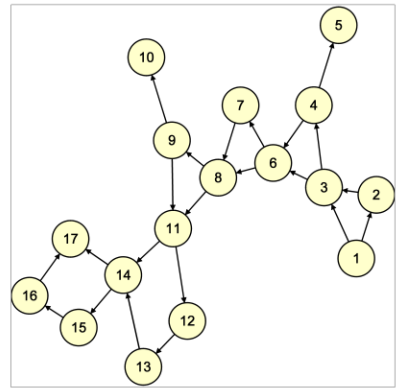

*Figure 2:* Elements II.5 *argument graph.*

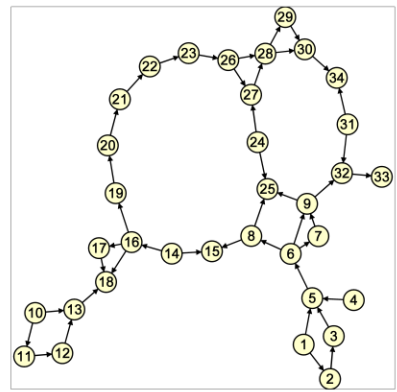

*Figure 3:* The Method I *argument graph.*

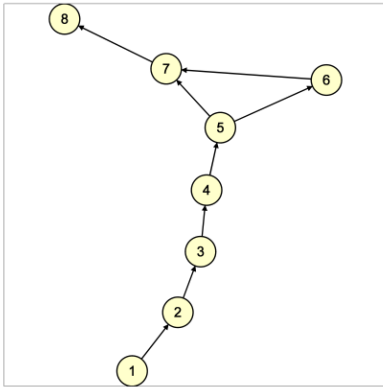

*Figure 4:* On Sphere and Cylinder 1.30 *argument graph.*

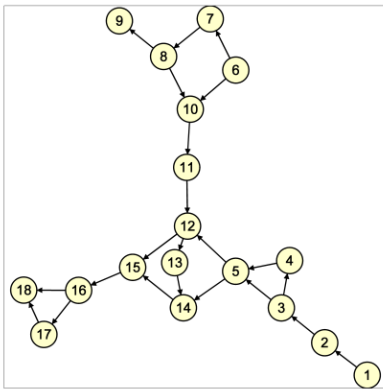

*Figure 5:* Spiral Lines 9 *argument graph.*

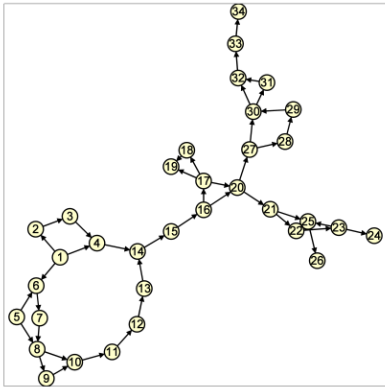

*Figure 6:* Conics 1.41 *argument graph.*

**QUANTITATIVE COMPARISON**

Having captured a representation of the artifacts as directed acyclic graphs, some things are self-evident, such as the differences in number of nodes. Indeed logical size as measured by number of assertions and arguments and related measures of logical complexity (Barany, 2009) represent cognitive reality (Netz, 1999, p. 200).

Here we focus on microstructure at the level of individual statements and arguments to see whether there are comparative patterns that match intuitions about cognitive process and epistemology.

One intuition we might have is that as compared to an individual intelligence, a collective intelligence might be able to incorporate many more ideas to argue for a new idea and that a single idea can be used to support a greater multitude of subsequent ideas in members of a collective. To test this mathematically, we look at the degree distributions of the arguments graphs. Nodes that are strong *hubs* (with high in-degree) would correspond to intense integration of information whereas nodes that are strong *authorities* (with high out-degree) would correspond to intense dissemination of information (Kleinberg, 1999). Figures 7 and 8 show the in-degree and out-degree distributions of the several argument graphs. As can be observed, the Polymath Project has a much broader degree distribution with stronger hubs and authorities. This suggests that the intuition was correct for these particular historical artifacts.

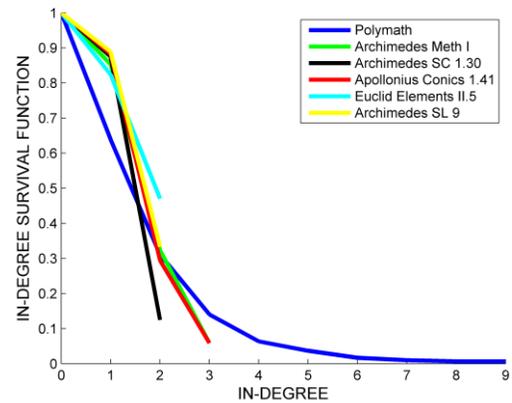

*Figure 7: In-degree survival functions for argument graphs.*

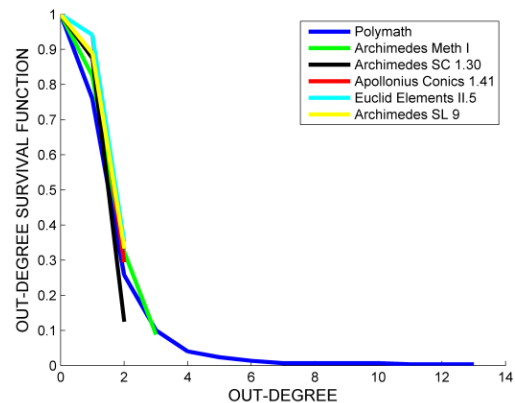

*Figure 8: Out-degree survival functions for argument graphs.*

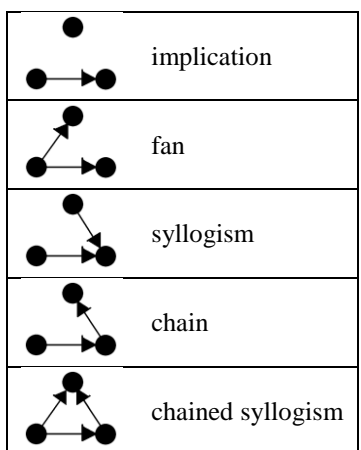

*Figure 9: Subgraphs.*

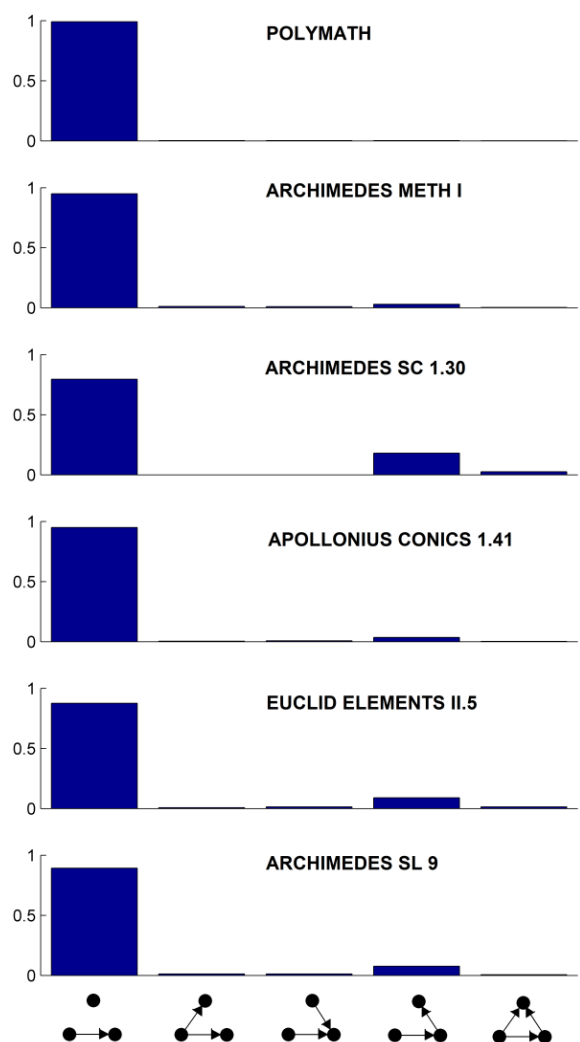

*Figure 10: Subgraph distributions for argument graphs.*

Next we consider subgraphs on three nodes, restricting ourselves to ones where there is at least one directed edge (so as to reduce the influence of the graph density in the analysis). Since argument graphs are acyclic, there are only five such subgraphs rather than fifteen. We use the names given in Figure 9 to denote them.

Figure 10 shows the subgraph distributions of the various argument graphs. In Polymath, the *implication* occurs with greater frequency than the other kinds of subgraphs, as compared to the Greek mathematicians. The *fan*, *chain*, and *syllogism* are particularly more frequent in Greek work.

When describing Greek work, Netz (1999, p. 199) had noted, "larger arguments exist, but are a rarity. We see therefore that arguments are short – and this is one way in which they are easy to follow." It seems from the subgraph analysis, however, that structurally speaking Polymath uses short and easy-to-follow arguments to an even greater extent than Greek mathematics from antiquity.

As noted previously, one common way to analyze subgraph distributions in network science is through the concept of structural motifs. Motifs are subgraphs overrepresented as compared to a null random graph ensemble (Milo et al., 2002). Subgraph distributions have also been used to define families of networks, such as the family of rate-limited information-processing networks (Milo et al., 2004). We leave investigation of motifs to future work due to the difficulty in defining an appropriate random graph ensemble of directed acyclic graphs. Similarly, the paucity of directed acyclic graphs in nature prevents a family characterization at this time.

To summarize, numerical analysis of degree and subgraph distributions has led to some insight into the cognitive process of D. H. J. Polymath and of several Greek mathematicians. As argued before, this cannot give universal insights into the difference between collective intelligence and individual intelligence. But as per our comparative cognitive history agenda, it does do so for two particular historical artifacts.

## **GOING FORWARD**

We have used two particular historical artifacts to try to tease out some potential differences between collective intelligence and individual intelligence in specific contexts. A systematic survey of many artifacts should provide even greater insight. Although not trying to fit universalist assumptions, one might nevertheless want to use a larger corpus to mitigate some concerns about confounding variables.

## **Limitations**

In the current case study, one kind of confounding variable is the time and place in history: historical artifacts are products of contextual milieus. Indeed the modern global mathematical enterprise is very different than the Greek mathematical enterprise of antiquity.

Another confound is the difference between final proof and the process of proving; between cognition in action and cognition in representation. We studied the final proof in the case of the ancient Greeks but the process of deduction in the case of Polymath. As noted by Gowers (2010), "although a proof, when written out, is a fairly linear object, starting from the premises and taking a direct route to the conclusion, the discovery of a proof is far from linear. It is more like a tree with many branches; but when you finally discover the branch that leads to the subbranch that leads to the twig that has at the end of it the solitary fruit that is the conclusion you were looking for, you throw away the rest of the tree."

A third is the branch of mathematics: one artifact was concerned with combinatorics whereas the other set of artifacts with geometry. Could the culture and cognitive process of deduction be different in these two subfields?

### Next Steps

More detailed mathematical analyses of artifacts beyond the basic results in the case study above may lead to enhanced understanding. Here, however, we highlight some ideas for limiting some confounding variables by the use of a larger corpus. With a large corpus of artifacts and their mathematically-defined features, data mining methods might even be able to discern common patterns across multifarious intelligences.

A first artifact to add is the structure of proof in the work of D. H. J. Polymath (Gowers, 2010) to supplement the artifact from the process of proving studied herein.

A second artifact to add is from the work by a modern mathematician. As an example, consider Terence Tao's recent work on localization and compactness properties of the Navier-Stokes global regularity problem. This result solves one of the Clay Mathematics Institute Millennium Prize Problems and is eligible for the $1 million prize. The implication diagram that explains the structure of argument is shown in Figure 11, (Tao, 2011).

A third artifact to add is the structure of proof in Archimedes' *Stomachion*, one of the earliest works in combinatorics rather than geometry in Greek antiquity (Netz and Noel, 2007).

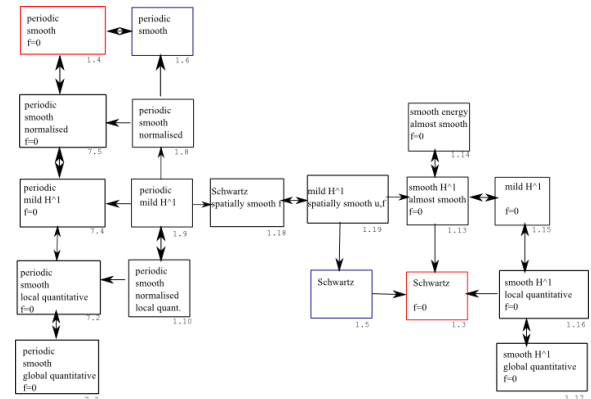

*Figure 11: Implication diagram for Terence Tao's work on the Navier-Stokes problem.*

### CONCLUSION

In this position paper, we put forth a research agenda to understand the 'doing of cognition' in collective intelligence through comparison with individual intelligence. To do so, we proposed the use of methods from cognitive history, with the attendant focus on understanding knowledge through its structure rather than its content by examination of specific cultural artifacts. We also demonstrated this with a small case study in the area of deductive reasoning.

A deeper analysis of mathematical deduction would be one way to step forward, however there are a whole host of other artifacts of collective intelligence that could be studied, cf. (Malone et al., 2010). As an example, a prominent use of collective intelligence is for calculation rather than proof; indeed calculation has just as deep a history as proof, with a multitude of cultural artifacts of individual intelligence (Narasimha, 2003).

Ultimately, comparative cognitive history may help shed light on the possibilities and constraints of new and emerging kinds of collective intelligence.

### ACKNOWLEDGEMENTS

Structural data on the Polymath Project was graciously provided by Justin Cranshaw and Aniket Kittur.